\documentclass{article_saj}
\usepackage{graphicx,saj,multicol,subeqnarray}
\usepackage{natbib}
\usepackage{float}
\usepackage{xcolor}
\usepackage{widetext}
\usepackage{url}
\usepackage{bm}
\usepackage{tikz} 
\usepackage{pifont} 
\usepackage{amsfonts}
\usepackage{amssymb}
\usepackage{amsmath,upgreek}
\usepackage{titlesec}
\titlelabel{\thetitle.\quad}
\definecolor{xlinkcolor}{cmyk}{1,0.6,0,0}
\usepackage[bookmarks=false,         
     pdfnewwindow=true,      
     colorlinks=true,    
     linkcolor=xlinkcolor,     
     citecolor=xlinkcolor,     
     filecolor=xlinkcolor,  
     urlcolor=xlinkcolor,      
final=true
]{hyperref}
\usepackage{subcaption}

\def\papertype{\ \hfill\ Preliminary report}

\def\udc{xx}
\hypersetup{draft}

\begin{document}

\parindent=.5cm
\baselineskip=3.8truemm
\columnsep=.5truecm
\newenvironment{lefteqnarray}{\arraycolsep=0pt\begin{eqnarray}}
{\end{eqnarray}\protect\aftergroup\ignorespaces}
\newenvironment{lefteqnarray*}{\arraycolsep=0pt\begin{eqnarray*}}
{\end{eqnarray*}\protect\aftergroup\ignorespaces}
\newenvironment{leftsubeqnarray}{\arraycolsep=0pt\begin{subeqnarray}}
{\end{subeqnarray}\protect\aftergroup\ignorespaces}
%


\markboth{\eightrm Proper motion of Cygnus loop shock filaments} 
{\eightrm M. Vu{\v c}eti{\' c} {\lowercase{\eightit{et al.}}}}

\begin{strip}

{\ }

\vskip-1cm

\publ

\type

{\ }


\title{Proper motion of Cygnus loop shock filaments}


\authors{M. Vu{\v c}eti{\' c}$^{1}$, N. Milanovi{\' c}$^{2}$, D. Uro{\v s}evi{\' c}$^{1}$, J. Raymond$^{3}$, D. Oni{\' c}$^{1}$, S. Milo{\v s}evi{\' c}$^{1}$ and N. Petrov$^{4}$}

\vskip3mm


\address{$^1$Department of Astronomy, Faculty of Mathematics,
University of Belgrade\break Studentski trg 16, 11000 Belgrade,
Serbia}


\Email{milica.vucetic@matf.bg.ac.rs}

\address{$^2$Max Planck Institute for Solar System Research, Justus-von-Liebig-Weg 3, 37077 G{\"o}ttingen, Germany}

\address{$^3$Harvard-Smithsonian Center for Astrophysics, 60 Garden Street, Cambridge, MA, 02138, USA}

\address{$^4$Institute of Astronomy and National Astronomical Observatory, Bulgarian Academy of Sciences, Tsarigradsko Shose 72, BG-1784 Sofia, Bulgaria}


\dates{May 18, 2020}{June 1, 2020}


    \summary{We determined shock speed in the Galactic supernova remnant Cygnus Loop, using proper motion of its optical filaments and the latest estimate for its distance. The proper motion was measured by comparing H$\alpha$ images of the remnant observed in two epochs: 1993 (Kitt Peak National Observatory) and 2018/2019 (National Astronomical Observatory Rozhen and Astronomical station Vidojevica). We derived shock speed for 35 {locations along different filaments}, which is twice as much as in earlier studies of north-eastern part of Cygnus Loop. For the first time, we have measured shock speed of radiative filaments in this region. Three of the analyzed {locations where we measured proper motion of filaments}  are radiative, based on their presence in [\hbox{S\,{\sc ii}}] images from the second epoch. The other filaments are non-radiative. The speed we obtained for the non-radiative filaments is in the range of 240--650 $\mathrm{km\ s^{-1}}$, with an estimate for the uncertainty of 70 $\mathrm{km\ s^{-1}}$. These values are mostly in agreement with previous studies. The  radiative filaments have  lower speed of 100--160$\pm${70} $\mathrm{km\ s^{-1}}$, which is in agreement with the assumption that they are older in evolutionary terms. This clear distinction between the speed of the two types of filaments proves that the [\hbox{S\,{\sc ii}}] emission can be used for identifying radiative filaments in supernova remnants.}



\end{strip}



\section{INTRODUCTION \label{S:intro}}

\indent
Cygnus Loop is a Galactic supernova remnant (SNR) G74.0-8.5, also known as the Veil nebula. Recently, its distance was estimated to 725$\pm$15 pc \cite[][]{Fesen2021}, based upon the parallaxes of stars available in the Gaia Early Data Release 3. With this result, the uncertainty of the Cygnus Loop distance was reduced down to a value comparable to its radius \cite[$\sim18$ pc;][]{Fesen2021}. This is significant, because the distances to the Galactic SNRs are not easy to determine. One way to estimate the distance is by taking into account distances to the stars known to be in front or behind the remnant. The ones that are behind can be recognized by absorption in their spectra, caused by the star's radiation passing through the SNR on its way to the observer. The estimated distance to Cygnus Loop ranged between 440 pc and 1400 pc over years \cite[see review by][]{Fesen2018a}. With the latest result by \citet{Fesen2021}, it eventually came close to the value of 770 pc given by \citet{Minkowski} more than 60 years ago.

This middle-aged remnant ($\sim$ 20000 years old) has an apparent diameter of $\sim$ 3$^{\circ}$ on the sky, and its size and brightness make it suitable for studying smaller structures (filaments) in the remnant. Newest findings suggest that Cygnus Loop's morphology is a consequence of its location in a low-density medium, far from the Galactic plane (b $\approx$ 8$^{\circ}$). There, it encounters interstellar clouds as it expands \citep{Fesen2018b}.
The brightest regions observed in the optical domain shape out nearly spherical western and eastern parts of the shell, whose radiation can be described by interaction of the shock wave with the surrounding dust clouds \citep{1986ApJ...300..675H}.

As the shock front in the SNR decelerates, the downstream region (behind the shock) loses more energy through radiation and cools down. This creates conditions in the downstream region suitable for formation of, for example, forbidden [\hbox{S\,{\sc ii}}] $\lambda\lambda$ 6716, 6731 emission lines  \citep{Raymond1979, 1997ApJS..112...49M}. There, the collisionally excited sulfur ions radiate strongly, typically producing line intensities that compare to H$\alpha$ as [\hbox{S\,{\sc ii}}]$/$H$\alpha>0.4$. Based on this, it is suggested that the filaments in the remnant that already entered the radiative phase of the evolution can be recognized through their [\hbox{S\,{\sc ii}}] emission.

Cygnus Loop contains both non-radiative and radiative filaments \citep{1994ApJ...420..721H}. Studies of north-eastern part of Cygnus Loop so far measured the speed of the non-radiative filaments using different techniques \cite[see][]{Salvesen, Medina, Sankrit2023}.  In this paper, for the first time, we determined the speed of the shock wave by measuring the proper motion of both non-radiative and radiative filaments in Cygnus Loop. The proper motion was measured by comparing H$\alpha$ observations from two epochs: 1993 and 2018/2019 \citep{2019Milanovic}. The speed of the filaments was derived using the latest estimate of the remnant's distance, mentioned above. In order to differentiate between radiative and non-radiative filaments, we observed Cygnus Loop through H$\alpha$ and [\hbox{S\,{\sc ii}}] narrowband filters. Radiative filaments of this SNR were recognized as those that are visible in images taken through narrowband [\hbox{S\,{\sc ii}}] filter \cite[like in e.g.][]{Vucetic}. In this way, we compared the speed of the radiative and the non-radiative filaments. We then tested the hypothesis that the radiative filaments should have lower speed, since they are supposed to be older in evolutionary terms.

The speed of the shocks producing the non-radiative filaments is important for other reasons as well. \cite{Salvesen} combined the shock speed with the Rankine-Hugoniot shock jump conditions to place upper limits on the fraction of shock energy that goes into cosmic rays.  More recently, \cite{2023Raymond} used the shock speed and jump conditions as one way to determine the electron-ion temperature ratios in shocks.

Optical data used in this paper are described in Section 2, while the method used for the filaments' proper motion measurements is presented in Section 3. Section 4 contains results and discussion. The paper summary is given in Section 5.

\begin{table*}
\caption{Log of observational data: instrument, date of the observations, exposure time $T_{\mathrm {exp}}$, size of the FOV, plate scale, used filters. Labels refer to the fields observed in the Epoch 2, marked in Fig.~\ref{fig1}}
\parbox{\textwidth}{
\vskip.25cm
\centerline{\begin{tabular}
{@{\extracolsep{0.0mm}}lcccccc@{}}
\hline
Instrument & Date & $T_{\mathrm{exp}}$ [s] & FOV & scale [$''$/px] & filters & labels \\
\hline
\hline
\multicolumn{7}{c}{\textbf{Epoch 1}} \\
KPNO & 1993 Aug 16 & 1500 & $0.7^\circ$ & 2.0268 & H$\alpha$ &  \\
\hline                                     
\multicolumn{7}{c}{\textbf{Epoch 2}} \\    
NAO Rozhen & 2018 Sept 12 & 1500-2700 & $5'\times 5'$ & 0.176 & H$\alpha$, [\hbox{S\,{\sc ii}}] & A - I \\
NAO Rozhen & 2019 Oct 1 & 2400 & $5'\times 5'$ & 0.176 & H$\alpha$ & K, L, M \\
ASV & 2019 Oct 28 & 3600 & $13'\times 13'$ & 0.39 & H$\alpha$, [\hbox{S\,{\sc ii}}]  & J \\
\hline
\end{tabular}}}
\label{table1}
\end{table*}


\begin{figure*}[h!]
  \begin{subfigure}{16cm}
    \centering\includegraphics[clip, trim={2cm 7cm 3cm 8cm}, width=\textwidth]{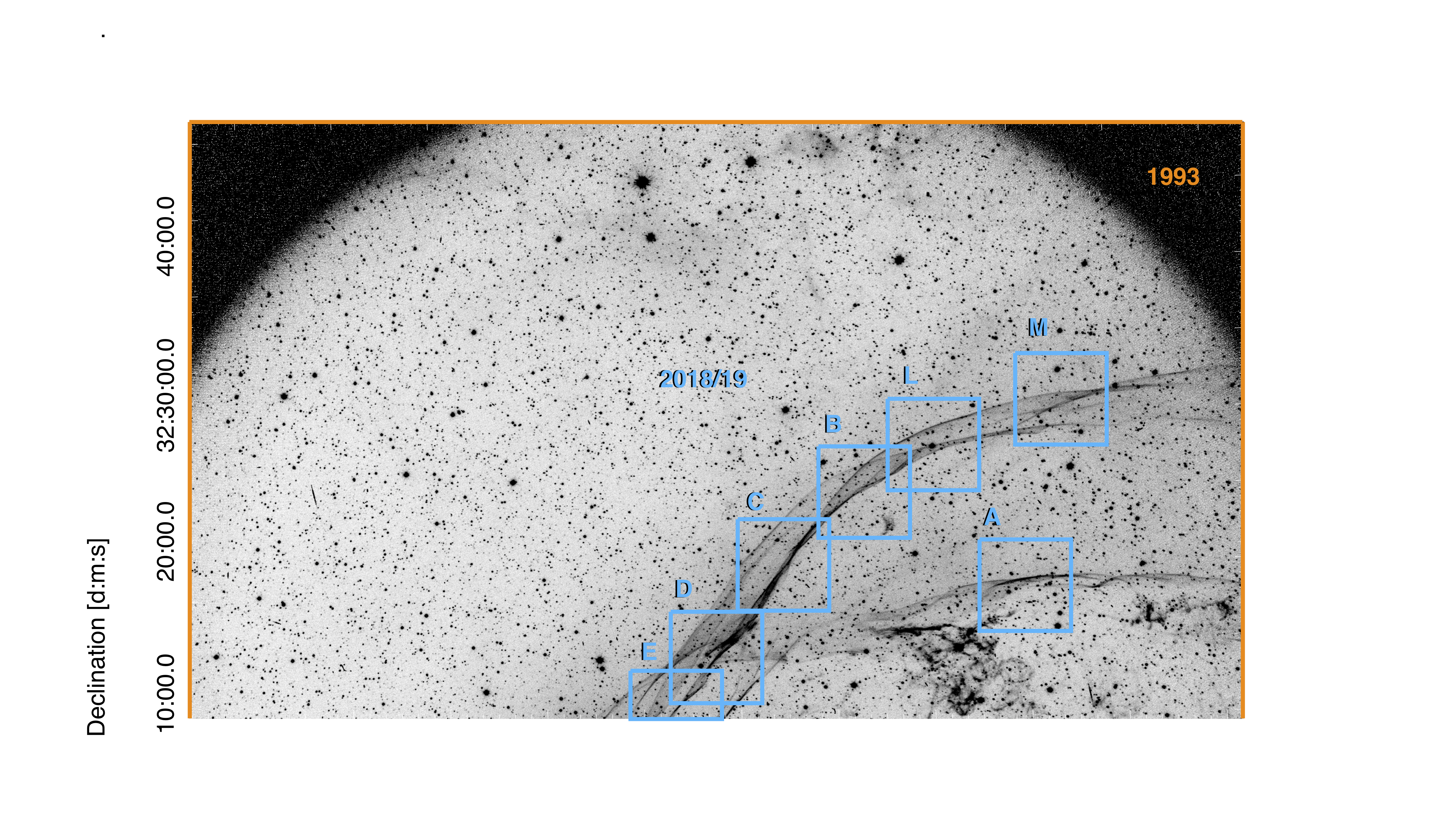}
  \end{subfigure}
  \begin{subfigure}{16cm}
    \centering\includegraphics[clip, trim={2cm 7cm 3cm 12cm}, width=\textwidth]{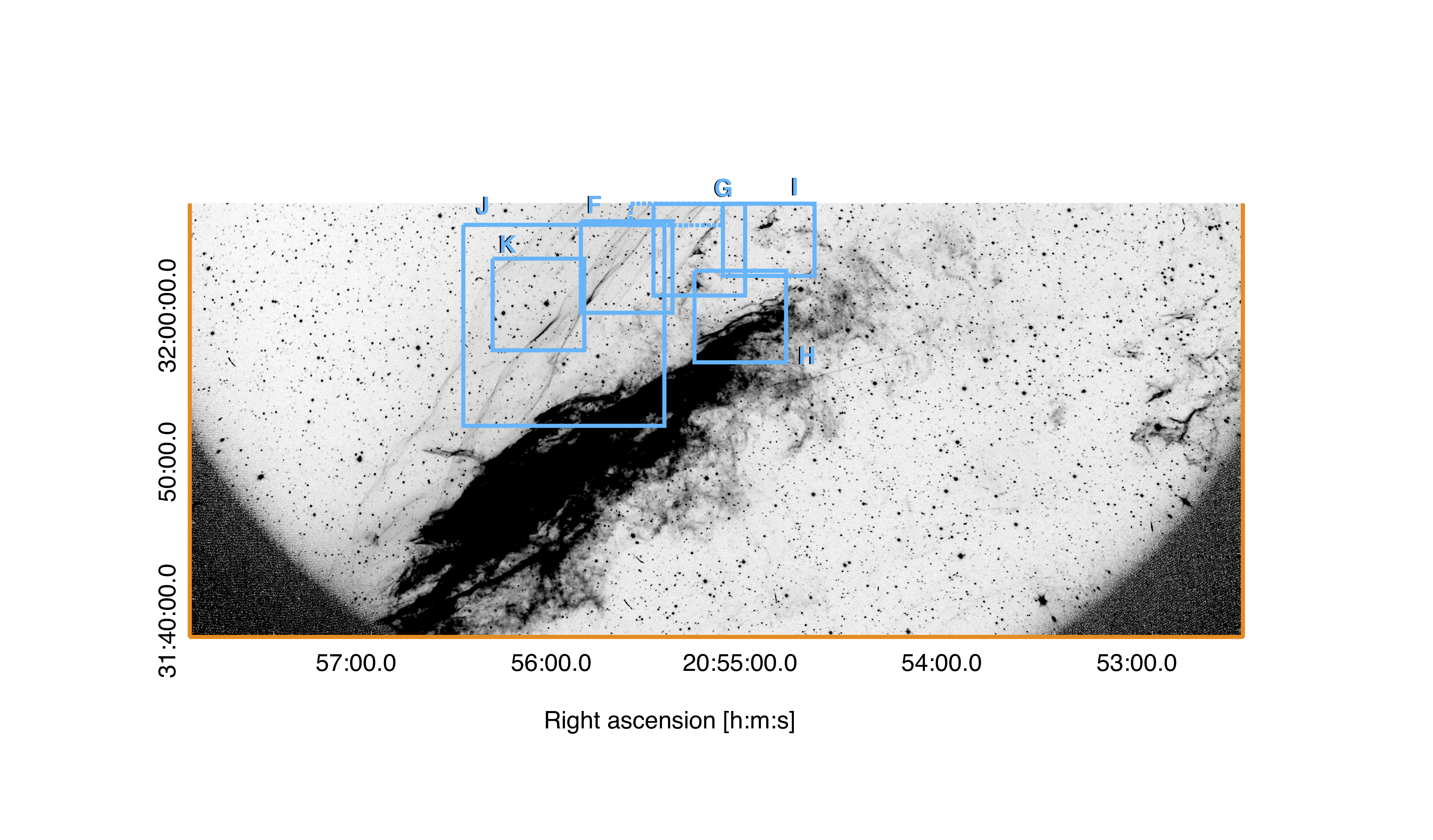}
  \end{subfigure}
  \caption{Context of our observations. The background image represents the Epoch 1 (year 1993) H$\alpha$ emission-line image of the north-eastern part of Cygnus Loop. The blue squares mark the fields observed in the Epoch 2 (years 2018, 2019), labeled with letters from A to M. Field E is split between the top and the bottom image from Epoch 1, and the dashed blue rectangle marks its bottom part. See Sect.~\ref{S:data}.}
  \label{fig1}
\end{figure*}

\section{OBSERVATIONS AND DATA REDUCTION \label{S:data}}

Shock speed in the Galactic SNR Cygnus Loop was determined using proper motion of its optical filaments. The proper motion was measured by comparing H$\alpha$ images of the remnant observed in two epochs: 1993 and 2018/2019. The speed of each filament was then calculated using the latest estimate for the distance to the object $d=725\pm$15 pc \citep{Fesen2021}.

Data from the Epoch 1 were acquired in 1993\footnote{We received these data in private communication with prof. Robert A. Fesen (Dartmouth College).}. The observations were performed using the Burrell-Schmidt telescope (diameter of 0.9 m) at the National Astronomical Observatory Kitt Peak (KPNO). The data consist of an image taken through a narrowband filter centered around the H$\alpha$ emission line, with the exposure time of 1500 s. The diameter of the field of view (FOV) is approximately 0.7$^{\circ}$, with a plate scale of $2.0268''$/px. These data are shown in Fig.~\ref{fig1}.

Data from the Epoch 2 were acquired in 2018 and 2019 at the National Astronomical Observatory Rozhen (NAO Rozhen) in Bulgaria, and in 2019 at the Astronomical Station Vidojevica (ASV) in Serbia. Observations from Epoch 2 consist of images taken through H$\alpha$ and [\hbox{S\,{\sc ii}}] narrowband filters. The H$\alpha$ images were used for measurements of the proper motion, while the [\hbox{S\,{\sc ii}}] images were used only for identification of the radiative filaments. 

At the NAO Rozhen, we used a 2 m  telescope with a FOV of $5'\times 5'$, and a plate scale of $0.176''$/px. Seeing conditions ranged from $1.2''$ to $1.8''$. In 2018 we observed 8 FOVs (locations marked with squares from A to I in Fig.~\ref{fig1}), and in 2019 we observed 3 FOVs (squares K, L, M in Fig.~\ref{fig1}). Our aim was to cover as many optical filaments as possible, that were contained in the Epoch 1 data. Each FOV was observed multiple times (5 -- 9, depending on FOV) with an exposure time of 300 s. Single exposures were combined in order to obtain higher signal-to-noise ratio. Narrowband H$\alpha$ and [\hbox{S\,{\sc ii}}] filters used from NAO Rozhen are $\sim$30{\AA} wide, and are centered at 6572{\AA} and 6719{\AA}, respectively.

Observations from the ASV were done using a 1.4 m telescope Milankovi{\' c}. This instrument has a larger FOV of $13'\times 13'$, and a plate scale of $0.39''$/px. There we observed one FOV, its location marked with J on Fig.~\ref{fig1}. Seeing conditions ranged from $1''$ to $2''$, and we took 12 images of 300 s through each filter. Narrowband H$\alpha$ and [\hbox{S\,{\sc ii}}] filters used from ASV are $\sim$50{\AA} wide, and are centered at 6561{\AA} and 6728{\AA}, respectively.

Details of observations in both epochs are given in Table \ref{table1}. We reduced the data from the Epoch 2 using standard procedures (bias and flat-field corrections, aligning and combining multiple exposures of the same FOV) in Python language. The data from the Epoch 1 we received as already reduced.

\section{FILAMENTS' PROPER MOTION MEASUREMENT \label{S:method}}

\begin{figure*}[h!]
\centering
\includegraphics[width=0.9\textwidth]
{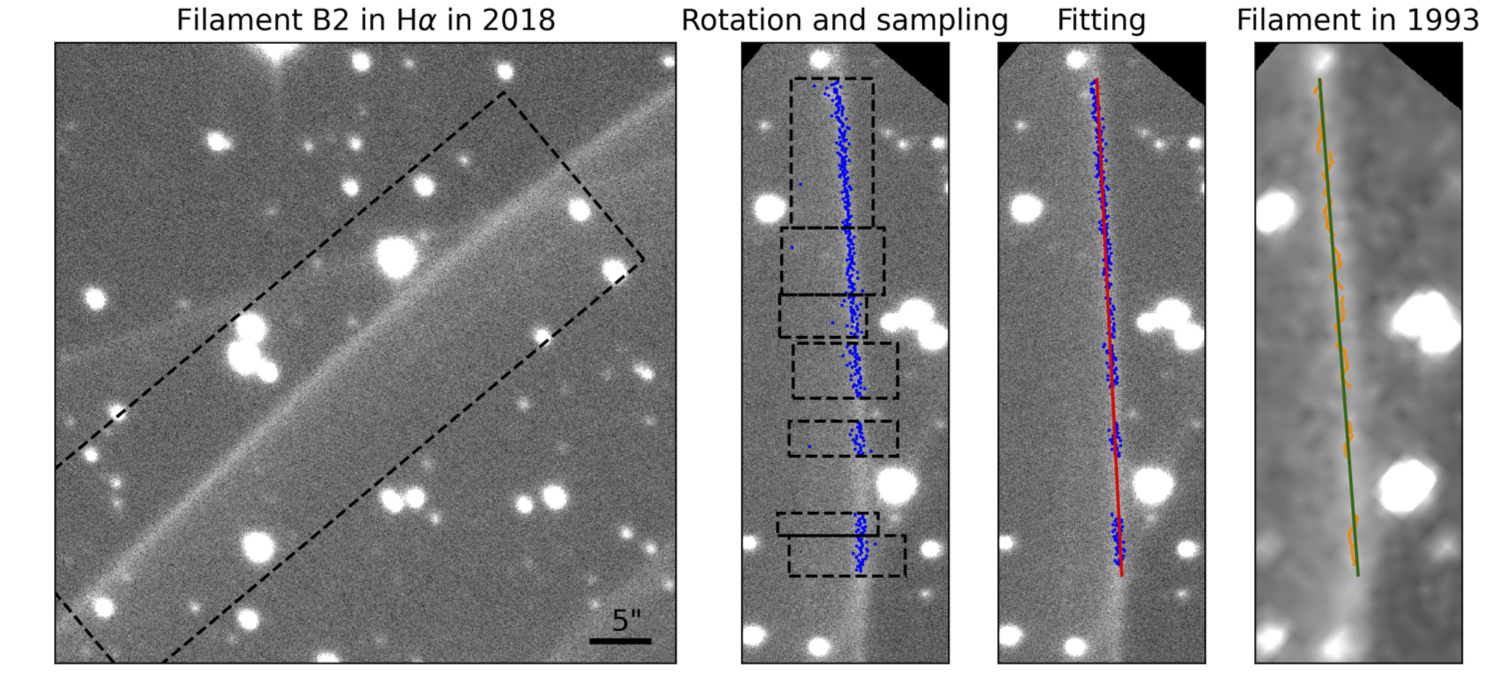}
\caption{Illustration of the measurement process for the filament B2. The first panel shows the filament, observed in the Epoch 2 in H$\alpha$ emission. The black rectangle outlines the area shown in the next two panels. In the second panel, the filament is rotated for $130^\circ$ clockwise, to be nearly vertical. The dashed rectangles show the slices selected for measurement. The points selected to represent the position of the filament (before filtering) are shown in blue. In the third panel, we show the filtered points and the linear fit to the filament (red). The fourth panel shows the same filament when observed in the Epoch 1, with the corresponding selected points and the linear fit. See Sect.~\ref{S:method}.}
\label{fig2}
\end{figure*}

To measure the shock speed of the optical filaments, we first linearly interpolated the image from the Epoch 1 to the plate scale of the Epoch 2 images ($2.0268''$/px to $0.176''$/px or $0.39''$/px). Then we aligned the Epoch 1 and Epoch 2 images according to the world coordinate system (WCS). For astrometric solution of the images (setting up the WCS grid and writing the corresponding FITS header keywords), we used the on-line platform http://nova.astrometry.net {\citep{2010Lang}. This web platform performs robust source detection in the input image, than asterisms - sets of four or five stars, are geometrically hashed and compared to the pre-indexed hashes, which are generated by using a combination of Gaia-DR2 and Tycho-2 catalogues.}

Next step was selecting suitable regions in both epochs for proper motion measurements. We searched for nearly linear filaments, without too many bright nearby stars, and without other close filaments that would interfere with the measurement. We also chose filaments that are bright enough and stand well above the background level, to make it easier for our routines to recognize the position of the filament.

In the measurement process, from each FOV marked in Fig.~\ref{fig1}, we selected smaller square subimages (size of $100'' \times 100'' $ in most cases), which contained one or more approximately linear filaments. Pairs of subimages at the same location were selected simultaneously from the Epoch 1 and 2, according to the WCS. In this way, we isolated a total of 43 filaments, for which we measured the proper motion using the method we discuss below. Some of these filaments were, however, excluded from the analysis after the measurement, for reasons that we will explain in Section 4.

Prior to the further measurement, each filament was rotated to be in a roughly vertical position. This rotation was done in order to measure proper motion in direction of $x$-axis on image. The angle of the rotation was estimated by eye for each filament, while the same angle was used for the filament in both epochs. Figure~\ref{fig2} shows an example of an approximately linear filament, cut from the region B, in Epochs 1 and 2. The filament is shown as it originally appears in the images, and after the rotation to make it nearly vertical.

After the rotation, in order to define the position of the filament in the subimages, we selected slices of the image that enclose the filament motion between the two epochs. These are the same slices in both epochs, marked with dashed rectangles in the second panel in Fig.~\ref{fig2}. The rectangles are placed in a way to cross the shock, but omit the areas containing stars, or parts of neighboring filaments. This method ensures that the brightness of the propagating shock will dominate its surroundings in the selected slices.

Further on, in each row of pixels in the slice, we identified the point with the maximum brightness and used it as the position of the filament in that row. Thus, the position of the filament in each epoch was described by one set of points, where the $y$ coordinate indicates the rows and the $x$ coordinate of the columns of pixels on the slice. These points are over-plotted in the second panel in Fig.~\ref{fig2}.

Although we carefully selected the slices suitable for measurement for each filament, sometimes the points that obviously belong to the surrounding stars were still caught in the sample. This was especially influenced by the shape of the stars in the Epoch 1 image, which appear larger and smeared because of the lower image quality. This trade-off between placing as large rectangles as possible (to increase the number of measured points) and avoiding stars made the measurement more difficult. Some of these points, that obviously do not belong to the filament, can be seen within the dashed rectangles in Fig.~\ref{fig2}.

Apart from the neighboring stars, points that belong to another nearby filament were sometimes caught in the sample, when there are two filaments crossing each other or generally being close in the image. Furthermore, in case of filaments that were not very bright, the selected points appeared more dispersed than for the brighter filaments. For all these reasons, after selecting the points that represent a filament, we applied a filtering process. For this, we used an iterative method of sigma-clipping, which we repeated three times. A straight line was fitted to the set of points, in each epoch. After the fitting, the measured points whose x-coordinate (in either of the two epochs) deviated by more than $\pm\ 2 \sigma$ from the linear fit were discarded. The points we show in the third panel in Fig.~\ref{fig2} are the final sample that represents the filament, after the filtering process.

Finally, after rotation and selection of the points that represent the filament in the images, we measured the proper motion of the filament. For this, we fitted the filament with a vertical regression $x=x(y)$ in each epoch (straight lines in the third and the fourth panel in Fig.~\ref{fig2}). Using the images from the two epochs, we calculated the difference between the fitted $x$ coordinates of the filament in each row. The mean value of these differences was then projected onto the direction perpendicular to the filament ($x$-axis), and taken as the proper motion of the filament over the time interval between the two epochs.

To estimate the expected value $p$ and the uncertainty $\Delta p$ of the proper motion, we used a bootstrap re-sampling procedure, where we repeated the above measuring process 10000 times. In each iteration, from the filtered set of points that represent the filament, a new random sample of points was chosen. The sample had always the same number of points as the initial set, with allowed repetitions of points. After all the iterations, we obtained a distribution of the measured proper motions, and we took its mean value as the final expected value of $p$. Since the dispersion of the distribution was always smaller than the pixel size, we used the pixel size ($0.176''$ or $0.39''$) for the uncertainty $\Delta p$.

\section{RESULTS AND DISCUSSION \label{S:results}}

\begin{figure*}[h!]
  \begin{subfigure}{8cm}
    \centering\includegraphics[clip, trim={26cm 2cm 28cm 7.5cm}, width=8.5cm]{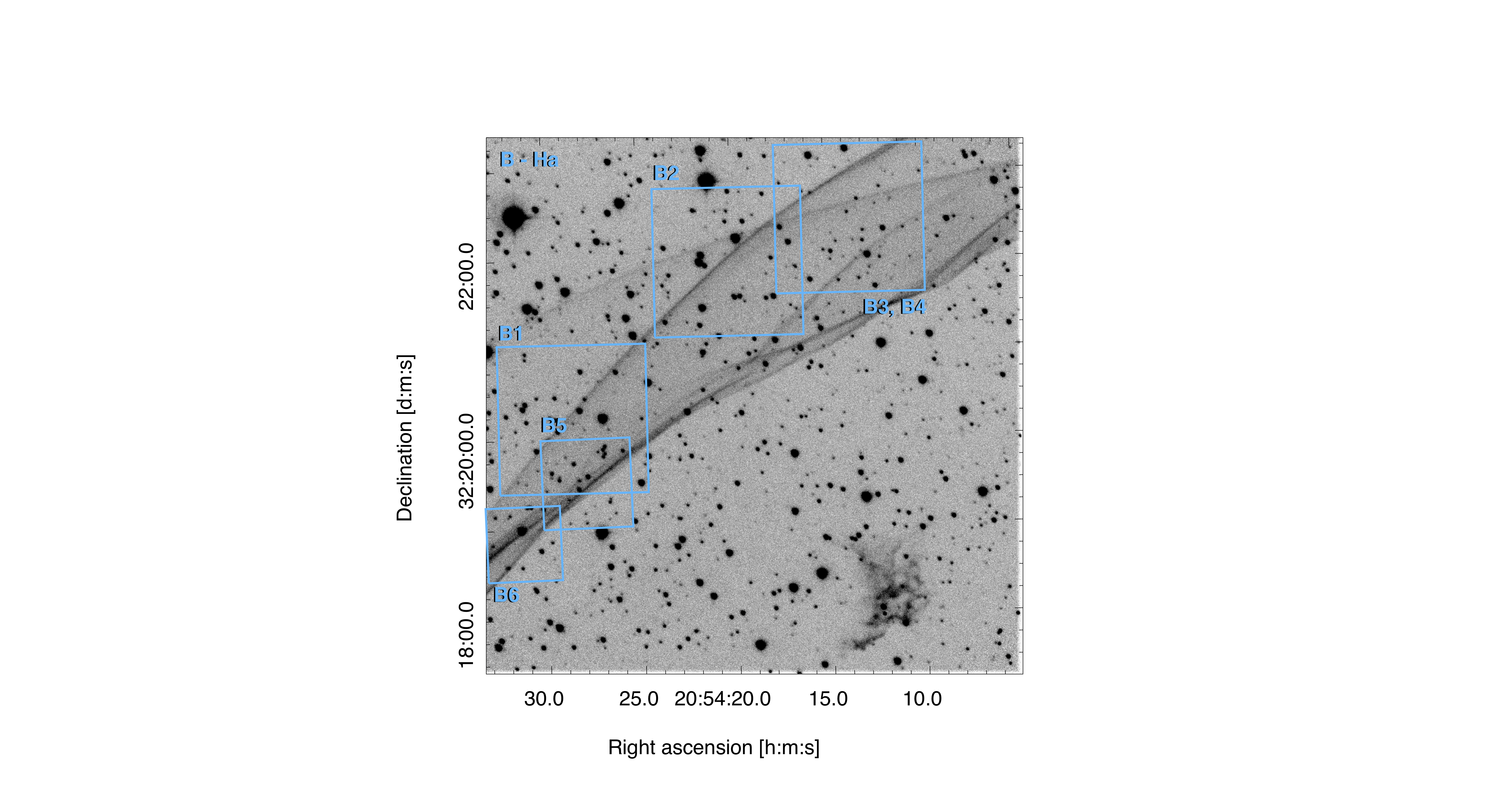}
  \end{subfigure}
  \begin{subfigure}{8cm}
    \centering\includegraphics[clip, trim={26cm 2cm 28cm 7.5cm}, width=8.5cm]{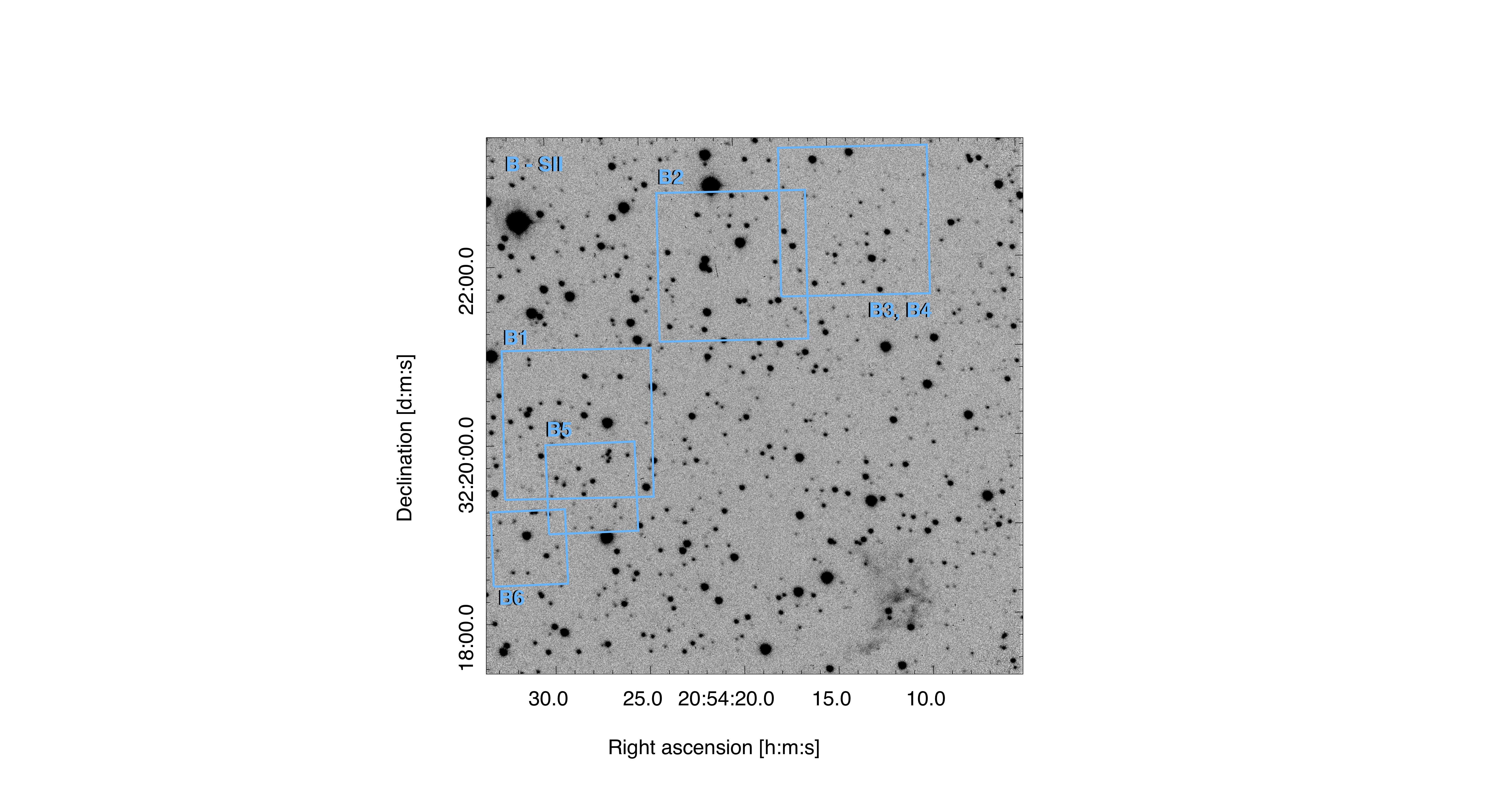}
  \end{subfigure}
  \begin{subfigure}{8cm}
    \centering\includegraphics[clip, trim={26cm 2cm 28cm 7.5cm}, width=8.5cm]{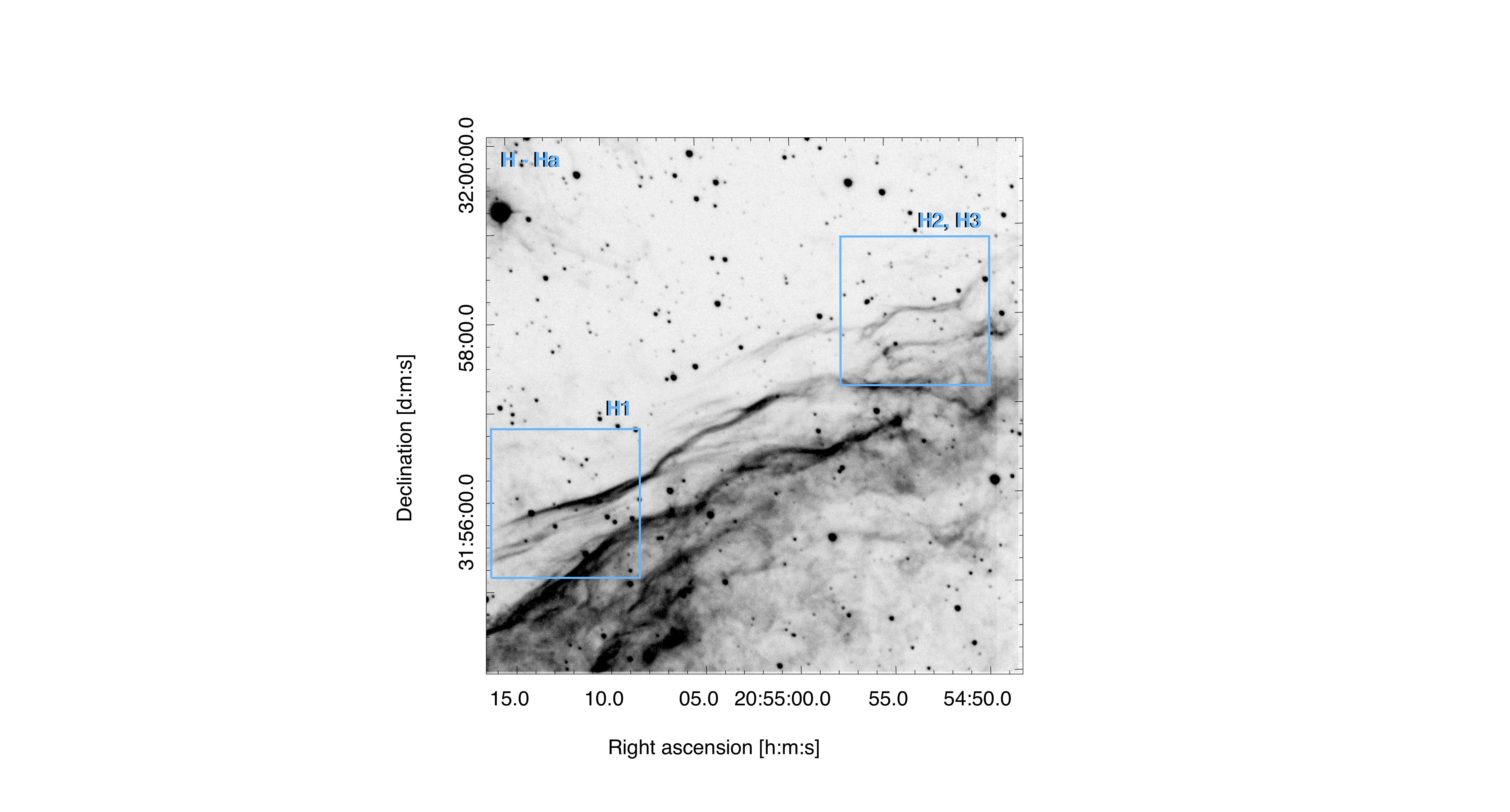}
  \end{subfigure}
  \begin{subfigure}{8cm}
    \centering\includegraphics[clip, trim={24cm 2cm 30cm 7.5cm}, width=8.5cm]{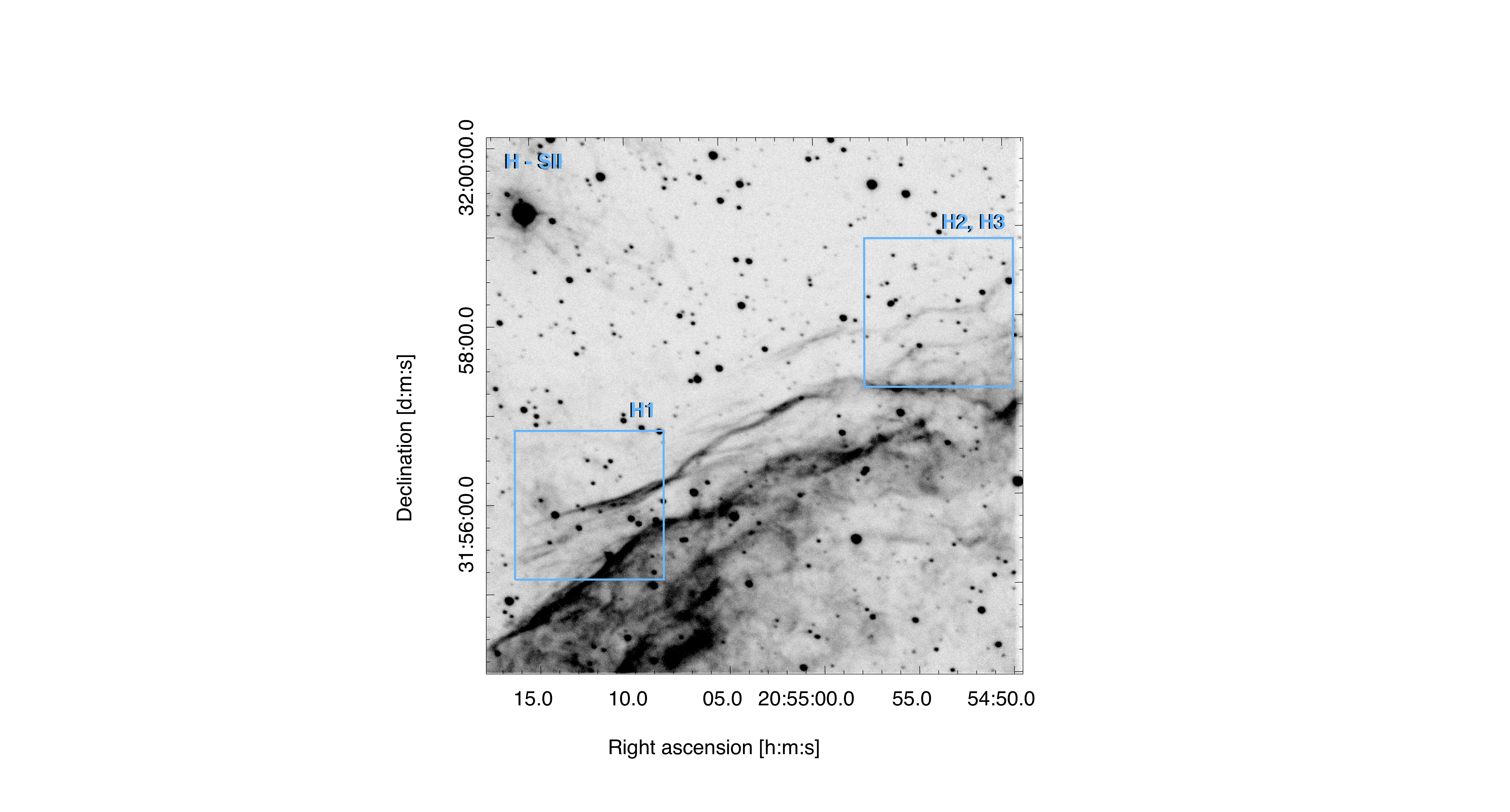}
  \end{subfigure}
  \caption{Top two images show H$\alpha$ and [\hbox{S\,{\sc ii}}] images of observed FOV $B$, which contains faster, non-radiative filaments, seen only on H$\alpha$ image. Bottom two images show FOV $H$, the only field of view where we see filaments emitting in [\hbox{S\,{\sc ii}}] lines, and for which we assume are in radiative phase of SNR evolution. For these filaments we measure the lowest velocities.}
  \label{fig3_added}
\end{figure*}

Assuming that the the shock front is moving in the plane of the sky and that the shock speed was constant between the two epochs of our observations, we used the measured proper motion of the filaments to derive their speed. In Table~\ref{table2} we list the coordinates, the measured proper motion and the derived speed $v$ for 35 {locations on different} filaments. Only {two} filaments, all located in our FOV $H$, were detected in [\hbox{S\,{\sc ii}}] images, and hence are designated as being in radiative phase. {Figure \ref{fig3_added} shows FOV $H$ and FOV $B$, to illustrate the difference between radiative and non-radiative filaments. Non-radiative filaments are not visible on [\hbox{S\,{\sc ii}}] images, while radiative filaments in FOV $H$ are clearly visible.} These filaments are  the most well-defined filaments in our sample, and we have measured their speed to be in range of 100--160$\pm$20 $\mathrm{km\ s^{-1}}$.

\begin{table*}
\caption{Measured filaments, their central coordinates, number of selected pixel rows, proper motion and speed. The filaments with a name in bold font, given at the end of the table are the ones visible in [\hbox{S\,{\sc ii}}] images. The measurements J1 and K1 represent the same filament, but observed with different instruments.}
\parbox{\textwidth}{
\vskip.25cm
\centerline{\begin{tabular}{@{\extracolsep{0.0mm}}lcccccc@{}}
  \hline
Filament	&	RA [h:m:s]	&	Dec [d:m:s]	&	number of rows	&	$p$ [$''$]	&	$v$ [$\mathrm{km\ s^{-1}}$] \\
 \hline
 \hline
A1	&	20:53:40.6	&	32:14:13.2	&	318	&	2.3	$\pm\ 0.6$  &	310 $\pm\ 70$	\\
A2	&	20:53:27.4	&	32:14:24.0	&	267	&	2.0	$\pm\ 0.6$	&	280 $\pm\ 70$	\\
B1	&	20:54:28.6	&	32:20:13.2	&	138	&	4.0	$\pm\ 0.6$	&	550 $\pm\ 70$	\\
B2	&	20:54:20.2	&	32:21:57.6	&	339	&	4.7	$\pm0.6$	&	650 $\pm\ 70$	\\
B3	&	20:54:13.7	&	32:22:26.4	&	181	&	3.6	$\pm0.6$	&	490 $\pm\ 70$	\\
B4	&	20:54:13.7	&	32:22:26.4	&	283	&	4.2	$\pm0.6$	&	580 $\pm\ 70$	\\
B5	&	20:54:27.9	&	32:19:30.0	&	207	&	2.0	$\pm0.6$	&	270 $\pm\ 70$	\\
B6	&	20:54:31.2	&	32:18:50.4	&	89	&	2.8	$\pm0.6$	&	380 $\pm\ 70$	\\
C1	&	20:54:51.1	&	32:17:20.4	&	399	&	2.5	$\pm0.6$	&	350 $\pm\ 70$	\\
C3	&	20:54:35.8	&	32:17:42.0	&	208	&	2.1	$\pm0.6$	&	290 $\pm\ 70$	\\
C4	&	20:54:44.7	&	32:15:46.8	&	269	&	3.5	$\pm0.6$	&	480 $\pm\ 70$	\\
C5	&	20:54:51.4	&	32:14:09.6	&	146	&	3.2	$\pm0.6$	&	440 $\pm\ 70$	\\
D1	&	20:54:57.6	&	32:08:09.6	&	191	&	2.5	$\pm0.6$	&	340 $\pm\ 70$	\\
D2	&	20:55:01.2	&	32:11:27.6	&	269	&	3.2	$\pm0.6$	&	440 $\pm\ 70$	\\
D3	&	20:55:05.0	&	32:10:15.6	&	151	&	2.2	$\pm0.6$	&	300 $\pm\ 70$	\\
E1	&	20:55:28.8	&	32:07:55.2	&	221	&	2.7	$\pm 0.6$	&	370 $\pm\ 70$	\\
E3	&	20:55:19.7	&	32:07:08.4	&	274	&	2.1	$\pm 0.6$	&	280 $\pm\ 70$	\\
E4	&	20:55:13.9	&	32:08:20.4	&	165	&	3.0	$\pm0.6$	&	400 $\pm\ 70$	\\
F1	&	20:55:34.6	&	32:01:44.4	&	227	&	2.1	$\pm0.6$	&	280 $\pm\ 70$	\\
F2	&	22:55:44.6	&	31:59:49.2	&	253	&	1.9	$\pm0.6$	&	260 $\pm\ 70$	\\
F3	&	20:55:42.7	&	32:02:13.2	&	264	&	3.3	$\pm0.6$	&	460 $\pm\ 70$	\\
G1	&	20:55:23.3	&	32:03:32.4	&	169	&	1.8	$\pm0.6$	&	240 $\pm\ 70$	\\
G2	&	20:55:15.1	&	32:03:32.4	&	211	&	2.3	$\pm0.6$	&	310 $\pm\ 70$	\\
G3	&	20:55:15.1	&	32:03:32.4	&	166	&	2.0	$\pm0.6$	&	280 $\pm\ 70$	\\
G4	&	20:55:19.7	&	32:02:38.4	&	167	&	2.0	$\pm0.6$	&	280 $\pm\ 70$	\\
G5	&	20:55:19.7	&	32:02:38.4	&	200	&	2.1	$\pm0.6$	&	290 $\pm\ 70$	\\
J1	&	20:55:55.4	&	31:56:20.4	&	146	&	2.4	$\pm0.6$	&	310 $\pm\ 70$	\\
J2	&	20:55:35.8	&	32:01:26.4	&	184	&	2.6	$\pm0.6$	&	340 $\pm\ 70$	\\
K1	&	20:55:55.4	&	31:56:20.4	&	335	&	2.5	$\pm0.6$	&	330 $\pm\ 70$	\\
L1	&	20:53:51.4	&	32:24:54.0	&	335	&	4.4	$\pm0.6$	&	580 $\pm\ 70$	\\
L2	&	20:53:59.5	&	32:24:21.6	&	371	&	4.3	$\pm0.6$	&	570 $\pm\ 70$	\\
M2	&	20:53:17.3	&	32:24:14.4	&	275	&	3.6	$\pm0.6$	&	470 $\pm\ 70$	\\
\textbf{H1}	&	{20:55:12.2}	&	{31:55:58.8}	&	{184}	&	{0.9	$\pm0.6$}	&	{130 $\pm\ 70$}	\\
\textbf{H2}	&	{20:54:53.0}	&	{31:57:57.6}	&	100	&	0.7	$\pm0.6$	&	100 $\pm\ 70$	\\
\textbf{H3}	&	{20:54:53.0}	&	{31:57:57.6}	&	86	&	1.1	$\pm0.6$	&	160 $\pm\ 70$	\\
\hline
\end{tabular}}}
\label{table2}
\end{table*}

\begin{figure*}[h!]
\centerline{\includegraphics[clip, trim={8cm 2cm 10cm 1cm}, width=\textwidth]{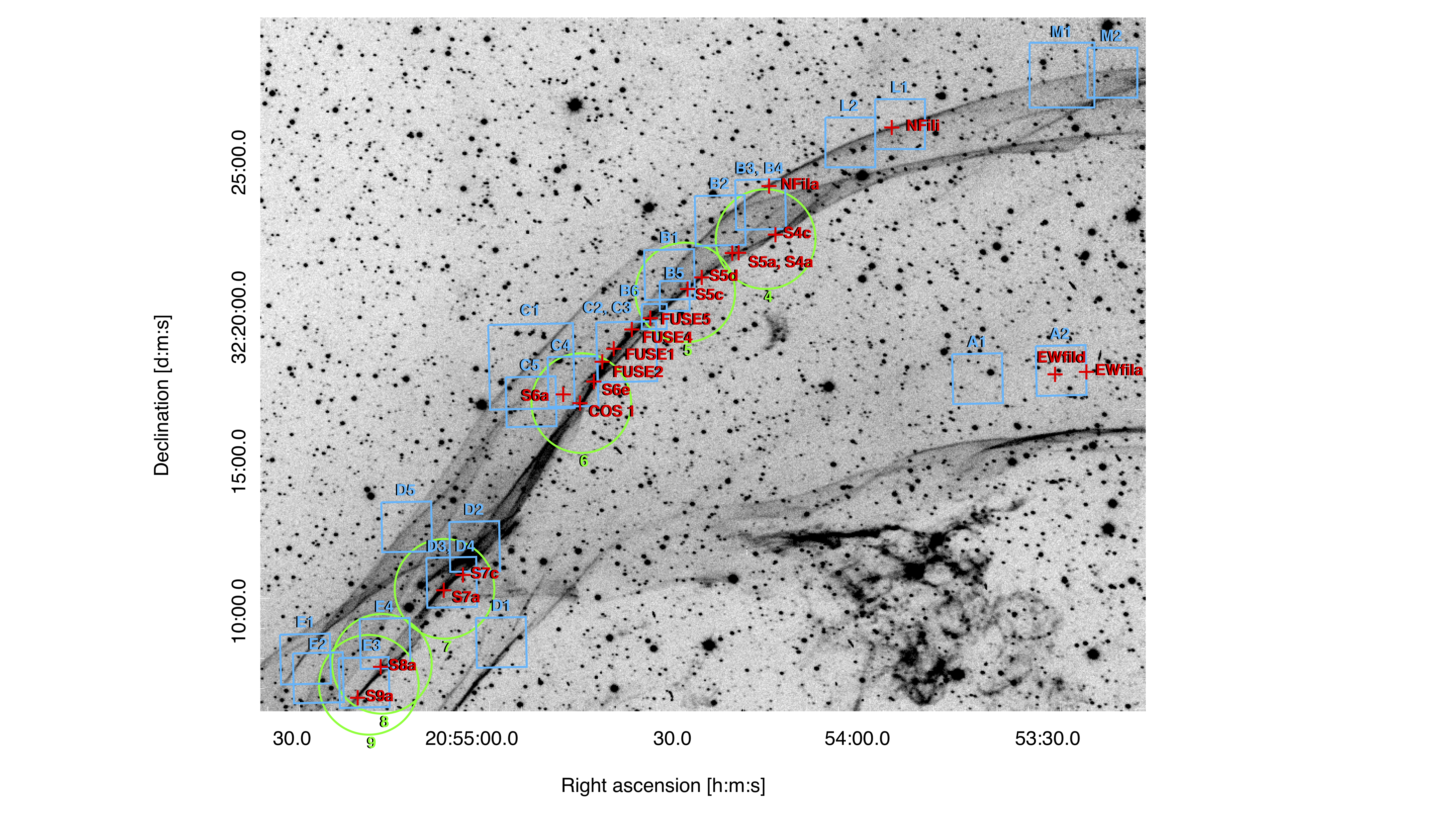}}
\caption{Comparison between the positions of measured filaments in this work and in the earlier studies. The image is located in the northern part of the general region we studied (Fig.~\ref{fig1}). The blue squares mark the filaments analyzed in this work. The green circles are the locations analyzed by \citet{Salvesen}, and the red crosses the ones analyzed by \citet{Medina}. See Sect.~\ref{S:results}.}
\label{fig3}
\end{figure*}

\begin{figure*}[h!]
\centerline{\includegraphics[clip, trim={11cm 13cm 7cm 13cm}, width=\textwidth]{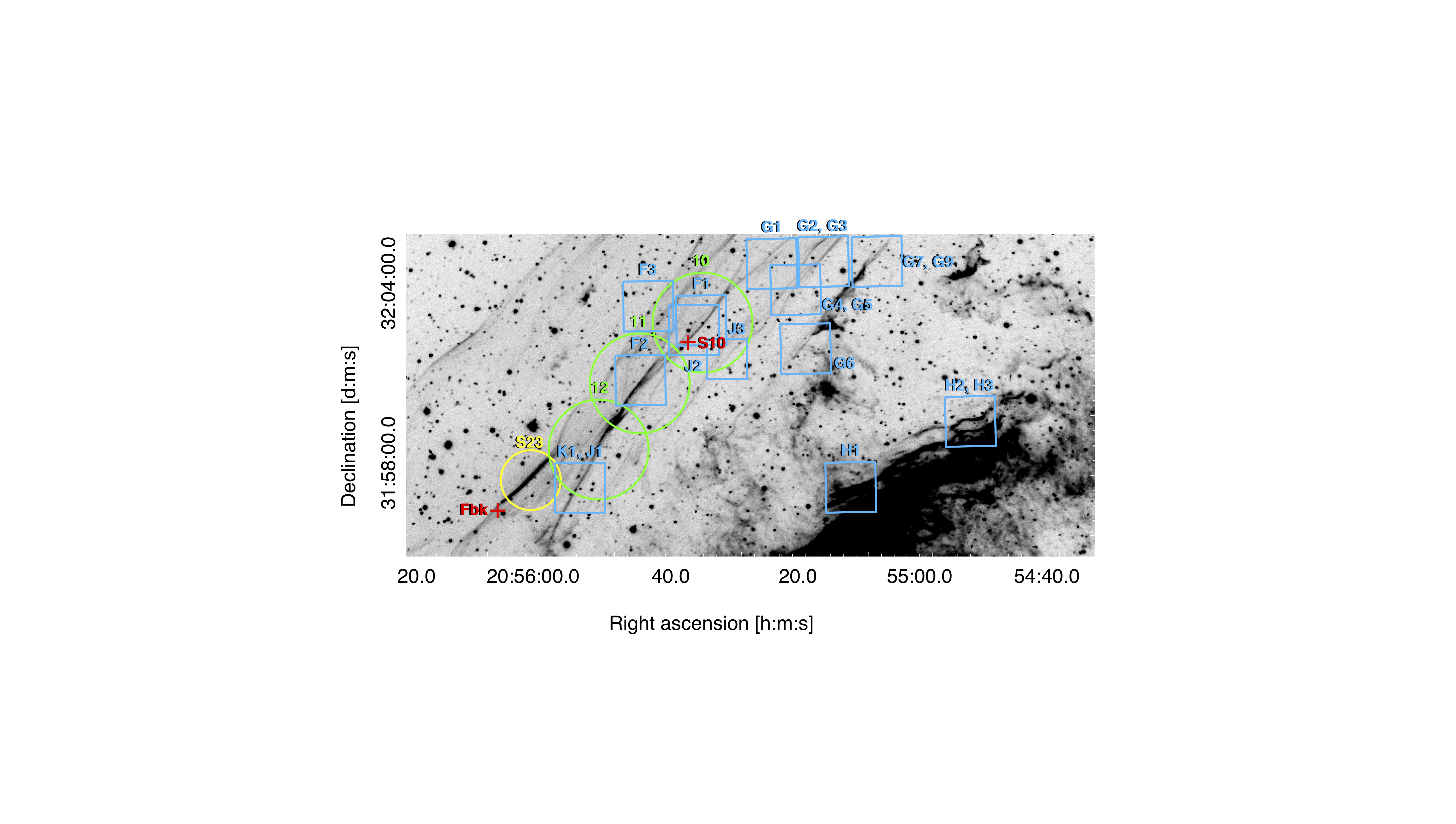}}

\caption{Comparison between the positions of measured filaments in this work and in the earlier studies. The image is located in the southern part of the general region we studied (Fig.~\ref{fig1}). The filaments are marked like in Fig.~\ref{fig3}, with an additional yellow circle highlighting the location analyzed by \citet{Sankrit2023}. See Sect.~\ref{S:results}.}
\label{fig4}
\end{figure*}

We estimated the uncertainty of the proper motion from the boot-strap measurement procedure to $0.2''$ -- $0.4''$.  Additional systematic uncertainty arises from astrometric calibration of the images from the both epochs. As mentioned in Section \ref{S:method}, astrometric calibration was made on astrometry.net web-platform. WCS uncertainty of Epoch 1 is calculated to $0.4''$, while WCS uncertainty of Epoch 2 is  $0.15''$. We calculate total error for proper motion $\Delta p$ as the sum in quadrature of errors of measurement procedure and WCS error, since they are independent. We calculated the uncertainty for the speed $\Delta v$ to be 70 $\mathrm{km\ s^{-1}}$ using the uncertainty for the proper motion $\Delta p=0.6''$ and  the distance to the remnant of 725 $\pm$ 15 pc \cite []{Fesen2021}. 

We compared our results to the measurements from earlier works of \citet{Salvesen}, \citet{Medina} and \citet{Sankrit2023}, who analyzed similar regions of Cygnus Loop like we did. 

In the work of \citet{Salvesen}, the speed of 18 non-radiative filaments was measured using a method similar to the one used in this work. The speed of the filaments was determined by comparing images from two different epochs (39.1 years apart). The filaments were also rotated prior to the measurements, so that they are nearly vertical, and the direction of the propagation of the shock wave is horizontal. After the rotation, the pixel values were summed along the vertical direction (columns), for each epoch individually, to obtain an intensity curve for the filament, whose peak should correspond to the brightest part of the shock wave front. Proper motion was then obtained from one-dimensional cross-correlation of the intensity curves in two epochs. In this paper, we also  tried  to use similar one-dimensional cross-correlation method, but it did not work, probably because of large difference in pixel size between epochs. 

In the work of \citet{Medina}, the authors performed spectroscopic measurements and obtained the intensity, width, and centroid of the narrow and the wide component of the H$\alpha$ line. The line widths of the broad component represent the post-shock kinetic energy and from shock models \citep{2001Ghavamian, 1980ApJ...235..186C} they estimated shock velocity. They measured 22 different positions in the remnant, located in non-radiative filaments. To improve the signal-to-noise ratio, some close positions were grouped. After comparing these measurements from spectroscopic observation with the model specified in the paper, the speed for 13 different positions (filaments) was obtained. It is important to note that this method of determining the speed of the filament does not depend on the distance, unlike the method used in \citet{Salvesen}, and in our work.

The most recent study of proper motion of Cygnus Loop filaments, based on HST H$\alpha$ images from three epochs \citep{Blair1999, Blair2005, Sankrit2023}, refers to the filament in our FOV $K$. Position of the filament measured by \cite{Sankrit2023} is marked in Fig.~\ref{fig4} as $S23$, and its median derived  velocity is 240 $\mathrm{km\ s^{-1}}$. For this particular filament, we could not perform proper motion measurement using our method, because the filament $S23$ is not linear.

In Table~\ref{table3}, we list the speed measured in the two mentioned studies \cite[see][]{Salvesen, Medina}, for the filaments that are close in position to the ones analyzed in this work. We mark these positions in Figs.~\ref{fig3} and~\ref{fig4}. In case of the study by \citet{Medina}, the positions marked with red crosses are the individual observation  obtained before grouping. Within the uncertainty range, most of the speeds obtained in this work are in agreement with the values for nearby filaments from the earlier studies. We note  that \cite{Salvesen} did not give any errors, while \cite{Medina} presented unrealistically small errors, since their shock speed errors  reflect only uncertainties in the H$\alpha$ line profile measurements. This could possibly explain  why these studies do not agree with each other for all positions. To this conclusion  we add that these studies used different methods for deriving shock velocities, as well as the fact that our method, same as methods used in \cite{Salvesen} and \cite{Sankrit2023} are actually measuring tangential speed, and hence represent a lower limit to the shock speed. 

We note that \cite{Salvesen} found that in some positions the shock speed was too low to account for the observed X-ray temperatures.  This discrepancy is partly resolved by using the \cite{Fesen2021} distance to the Cygnus Loop to convert proper motion to shock speed, rather than the smaller 561$\pm$61 pc distance used by \cite{Salvesen}.  However, as noted by \cite{Medina}, the problem still remains for some filaments. In some cases (e.g., our filament C4) our new proper motion values resolve the problem, but not in all cases. The most likely explanation is that we measure the proper motions of individual sharp filaments, while the measured X-ray temperatures correspond to averages along the line of sight.

\begin{table*}
\caption{Comparison between the speed of the filaments obtained in this work and in the studies by \citet{Salvesen} and \citet{Medina}. The speed is given in $\mathrm{km\ s^{-1}}$. The speed $v_{\mathrm{S,corr}}$	was calculated by adjusting the original speed from their study ($v_{\mathrm{S}}$) with the latest value for the remnant's distance \cite[725$\pm$15 pc;][]{Fesen2021}.}
\parbox{\textwidth}{
\vskip.25cm
\centerline{\begin{tabular}{@{\extracolsep{0.0mm}}lc|ccc|cc@{}}
  \hline
 \multicolumn{2}{c}{This work}	&	 \multicolumn{3}{|c|}{ Salvesen et al. (2009)}	&	\multicolumn{2}{c}{ Medina et al. (2014)}	\\
Filament	&	$v$	&	Filament 	&	$v_{\mathrm{S}}$	&	$v_{\mathrm{S,corr}}$	&	Filament	&	$v_\mathrm{M}$	\\
  \hline
  \hline
 A2 & 280 $\pm\ 70$ &    &     &     &  EWfil (EWfild) & 380 $\pm\ 4$ \\
 B4 & 580 $\pm\ 70$ &    &     &     &    NFil (NFila) & 429 $\pm\ 22$ \\
 B5 & 270 $\pm\ 70$ &  5 & 278 & 316 &        S5 (S5c) & 376 $\pm\ 4$ \\
 B6 & 380 $\pm\ 70$ &    &     &     &   FUSEB (FUSE5) & 414 $\pm\ 5$ \\
 C3 & 290 $\pm\ 70$ &    &     &     &   FUSEB (FUSE4) & 414 $\pm\ 5$ \\
 C4 & 480 $\pm\ 70$ &  6 & 333 & 379 & S6 (S6a), COS 1 & 391 $\pm\ 3$, 405  $\pm\ 8$\\
 D3 & 300 $\pm\ 70$ &  7 & 240 & 273 &        S7 (S7a) & 363 $\pm\ 5$ \\
 E3 & 280 $\pm\ 70$ &  9 & 294 & 335 &       S89 (S9a) & 457 $\pm\ 22$, 382 $\pm\ 6$ \\
 F1 & 280 $\pm\ 70$ & 10 & 279 & 318 &             S10 & 342 $\pm\ 46$ \\
 F2 & 260 $\pm\ 70$ & 11 & 254 & 289 &                 &               \\
 J2 & 340 $\pm\ 70$ & 10 & 279 & 318 &             S10 & 342 $\pm\ 46$ \\
 L1 & 580 $\pm\ 70$ &    &     &     &    Nfili (Nfil) & 429 $\pm\ 22$ \\
\hline
\end{tabular}}}
\label{table3}
\end{table*}

\section{CONCLUSIONS \label{S:concl}}

We studied the shock speed in the galactic supernova remnant Cygnus Loop, using the narrowband H$\alpha$ images from two different epochs. We obtained the speed by measuring the proper motion of the filaments in the remnant, and using the latest estimate for the remnant's distance.

In this work, we analyzed a total of 35 filaments, which is twice as much as in the earlier studies. We estimated their speed, including both non-radiative and radiative filaments. The radiative filaments were selected based on their visibility in [\hbox{S\,{\sc ii}}] images from the second epoch. Within our sample, there were  three radiative filaments. 

The speed we obtained for the non-radiative filaments, those filament which were not detected in [\hbox{S\,{\sc ii}}] images, is in the range of  240--650 $\mathrm{km\ s^{-1}}$ with uncertainty of 70 $\mathrm{km\ s^{-1}}$. These values are mostly in agreement with the earlier studies.  The two radiative filaments, which are probably the most well-defined filaments in our sample, have a speed of 100--160$\pm$70 $\mathrm{km\ s^{-1}}$. These values are  lower than for the non-radiative filaments, and are in agreement with the expected evolution scheme of SNRs. Speeds for the radiative filaments derived in this work are similar to those derived from HST proper motions in the bright western filaments of the Cygnus Loop by \cite{2020ApJ...903....2R}. Our results confirm that [\hbox{S\,{\sc ii}}] emission can be used for recognizing the radiative phase of the SNR evolution.

Our intention for future work is to obtain radio observations of confirmed radiative and non-radiative filaments in Cygnus Loop, in order to search for possible differences in the radio spectral index slopes of these two types of filaments. By this, we  want to study  the particle acceleration processes in different types of shocks.


\acknowledgements{Authors would like to thank anonymous referee for valuable suggestions which improved this paper. Authors also thank prof. Robert A. Fesen from Dartmouth College for sharing his images from year 1993 with us. M.V., D.U., D.O. and S.M. acknowledge funding provided by the University of Belgrade - Faculty of Mathematics (the contract \textnumero451-03-47/2023-01/200104) through the grant by the Ministry of Science, Technological Development and Innovation of the Republic of Serbia. M.V., D.U., N.P., D.O. and S.M. also acknowledge the observing and financial grant support from the Institute of Astronomy and NAO Rozhen through the bilateral joint research project between Bulgarian Academy of Sciences and Serbian Academy of Sciences and Arts. Authors acknowledge support by the Astronomical Station Vidojevica and funding from the Ministry of science, technological development and innovation of the Republic of Serbia, contract No. 451-03-47/2023-01/200002 and by the EC through project BELISSIMA (call FP7-REGPOT-2010-5, No. 265772). }



\newcommand\eprint{in press }

\bibsep=0pt

\bibliographystyle{aa_url_saj}

{\small

\bibliography{cygloop_saj}

\begin{thebibliography}{20}
\expandafter\ifx\csname natexlab\endcsname\relax\def\natexlab#1{#1}\fi

\bibitem[{{Blair} {et~al.}(2005){Blair}, {Sankrit}, \& {Raymond}}]{Blair2005}
{Blair}, W.~P., {Sankrit}, R., and {Raymond}, J.~C. 2005,
  \href{https://ui.adsabs.harvard.edu/abs/2005AJ....129.2268B}{\aj},
  \href{https://ui.adsabs.harvard.edu/abs/2005AJ....129.2268B}{129, 2268}

\bibitem[{{Blair} {et~al.}(1999){Blair}, {Sankrit}, {Raymond}, \&
  {Long}}]{Blair1999}
{Blair}, W.~P., {Sankrit}, R., {Raymond}, J.~C., and {Long}, K.~S. 1999,
  \href{https://ui.adsabs.harvard.edu/abs/1999AJ....118..942B}{\aj},
  \href{https://ui.adsabs.harvard.edu/abs/1999AJ....118..942B}{118, 942}

\bibitem[{{Chevalier} {et~al.}(1980){Chevalier}, {Kirshner}, \&
  {Raymond}}]{1980ApJ...235..186C}
{Chevalier}, R.~A., {Kirshner}, R.~P., and {Raymond}, J.~C. 1980,
  \href{https://ui.adsabs.harvard.edu/abs/1980ApJ...235..186C}{\apj},
  \href{https://ui.adsabs.harvard.edu/abs/1980ApJ...235..186C}{235, 186}

\bibitem[{{Fesen} {et~al.}(2018{\natexlab{a}}){Fesen}, {Neustadt}, {Black}, \&
  {Milisavljevic}}]{Fesen2018a}
{Fesen}, R.~A., {Neustadt}, J. M.~M., {Black}, C.~S., and {Milisavljevic}, D.
  2018{\natexlab{a}},
  \href{https://ui.adsabs.harvard.edu/abs/2018MNRAS.475.3996F}{\mnras},
  \href{https://ui.adsabs.harvard.edu/abs/2018MNRAS.475.3996F}{475, 3996}

\bibitem[{{Fesen} {et~al.}(2021){Fesen}, {Weil}, {Cisneros}, {Blair}, \&
  {Raymond}}]{Fesen2021}
{Fesen}, R.~A., {Weil}, K.~E., {Cisneros}, I., {Blair}, W.~P., and {Raymond},
  J.~C. 2021,
  \href{https://ui.adsabs.harvard.edu/abs/2021MNRAS.507..244F}{\mnras},
  \href{https://ui.adsabs.harvard.edu/abs/2021MNRAS.507..244F}{507, 244}

\bibitem[{{Fesen} {et~al.}(2018{\natexlab{b}}){Fesen}, {Weil}, {Cisneros},
  {Blair}, \& {Raymond}}]{Fesen2018b}
{Fesen}, R.~A., {Weil}, K.~E., {Cisneros}, I.~A., {Blair}, W.~P., and
  {Raymond}, J.~C. 2018{\natexlab{b}},
  \href{https://ui.adsabs.harvard.edu/abs/2018MNRAS.481.1786F}{\mnras},
  \href{https://ui.adsabs.harvard.edu/abs/2018MNRAS.481.1786F}{481, 1786}

\bibitem[{{Ghavamian} {et~al.}(2001){Ghavamian}, {Raymond}, {Smith}, \&
  {Hartigan}}]{2001Ghavamian}
{Ghavamian}, P., {Raymond}, J., {Smith}, R.~C., and {Hartigan}, P. 2001,
  \href{https://ui.adsabs.harvard.edu/abs/2001ApJ...547..995G}{\apj},
  \href{https://ui.adsabs.harvard.edu/abs/2001ApJ...547..995G}{547, 995}

\bibitem[{{Hester} and {Cox}(1986){Hester} \& {Cox}}]{1986ApJ...300..675H}
{Hester}, J.~J. and {Cox}, D.~P. 1986,
  \href{https://ui.adsabs.harvard.edu/abs/1986ApJ...300..675H}{\apj},
  \href{https://ui.adsabs.harvard.edu/abs/1986ApJ...300..675H}{300, 675}

\bibitem[{{Hester} {et~al.}(1994){Hester}, {Raymond}, \&
  {Blair}}]{1994ApJ...420..721H}
{Hester}, J.~J., {Raymond}, J.~C., and {Blair}, W.~P. 1994,
  \href{https://ui.adsabs.harvard.edu/abs/1994ApJ...420..721H}{\apj},
  \href{https://ui.adsabs.harvard.edu/abs/1994ApJ...420..721H}{420, 721}

\bibitem[{{Lang} {et~al.}(2010){Lang}, {Hogg}, {Mierle}, {Blanton}, \&
  {Roweis}}]{2010Lang}
{Lang}, D., {Hogg}, D.~W., {Mierle}, K., {Blanton}, M., and {Roweis}, S. 2010,
  \href{https://ui.adsabs.harvard.edu/abs/2010AJ....139.1782L}{\aj},
  \href{https://ui.adsabs.harvard.edu/abs/2010AJ....139.1782L}{139, 1782}

\bibitem[{{Matonick} and {Fesen}(1997){Matonick} \&
  {Fesen}}]{1997ApJS..112...49M}
{Matonick}, D.~M. and {Fesen}, R.~A. 1997,
  \href{https://ui.adsabs.harvard.edu/abs/1997ApJS..112...49M}{\apjs},
  \href{https://ui.adsabs.harvard.edu/abs/1997ApJS..112...49M}{112, 49}

\bibitem[{Medina {et~al.}(2014)Medina, Raymond, Edgar, Caldwell, Fesen, \&
  Milisavljevic}]{Medina}
Medina, A.~A., Raymond, J.~C., Edgar, R.~J., {et~al.} 2014, The Astrophysical
  Journal, 791, 30

\bibitem[{{Milanovic} {et~al.}(2019){Milanovic}, {Vucetic}, {Onic}, {Raymond},
  \& {Urovsevic}}]{2019Milanovic}
{Milanovic}, N., {Vucetic}, M., {Onic}, D., {Raymond}, J., and {Urovsevic}, D.
  2019, in Supernova Remnants: An Odyssey in Space after Stellar Death II, 127

\bibitem[{{Minkowski}(1958)}]{Minkowski}
{Minkowski}, R. 1958,
  \href{https://ui.adsabs.harvard.edu/abs/1958RvMP...30.1048M}{Reviews of
  Modern Physics},
  \href{https://ui.adsabs.harvard.edu/abs/1958RvMP...30.1048M}{30, 1048}

\bibitem[{{Raymond}(1979)}]{Raymond1979}
{Raymond}, J.~C. 1979,
  \href{https://ui.adsabs.harvard.edu/abs/1979ApJS...39....1R}{\apjs},
  \href{https://ui.adsabs.harvard.edu/abs/1979ApJS...39....1R}{39, 1}

\bibitem[{{Raymond} {et~al.}(2023){Raymond}, {Ghavamian}, {Bohdan}, {Ryu},
  {Niemiec}, {Sironi}, {Tran}, {Amato}, {Hoshino}, {Pohl}, {Amano}, \&
  {Fiuza}}]{2023Raymond}
{Raymond}, J.~C., {Ghavamian}, P., {Bohdan}, A., {et~al.} 2023,
  \href{https://ui.adsabs.harvard.edu/abs/2023ApJ...949...50R}{\apj},
  \href{https://ui.adsabs.harvard.edu/abs/2023ApJ...949...50R}{949, 50}

\bibitem[{{Raymond} {et~al.}(2020){Raymond}, {Slavin}, {Blair}, {Chilingarian},
  {Burkhart}, \& {Sankrit}}]{2020ApJ...903....2R}
{Raymond}, J.~C., {Slavin}, J.~D., {Blair}, W.~P., {et~al.} 2020,
  \href{https://ui.adsabs.harvard.edu/abs/2020ApJ...903....2R}{\apj},
  \href{https://ui.adsabs.harvard.edu/abs/2020ApJ...903....2R}{903, 2}

\bibitem[{Salvesen {et~al.}(2009)Salvesen, Raymond, \& Edgar}]{Salvesen}
Salvesen, G., Raymond, J.~C., and Edgar, R.~J. 2009, The Astrophysical Journal,
  702, 327–339

\bibitem[{{Sankrit} {et~al.}(2023){Sankrit}, {Blair}, \&
  {Raymond}}]{Sankrit2023}
{Sankrit}, R., {Blair}, W.~P., and {Raymond}, J.~C. 2023,
  \href{https://ui.adsabs.harvard.edu/abs/2023ApJ...948...97S}{\apj},
  \href{https://ui.adsabs.harvard.edu/abs/2023ApJ...948...97S}{948, 97}

\bibitem[{{Vu{\v{c}}eti{\'c}} {et~al.}(2015){Vu{\v{c}}eti{\'c}},
  {{\'C}iprijanovi{\'c}}, {Pavlovi{\'c}}, {Pannuti}, {Petrov}, {G{\"o}ker}, \&
  {Ercan}}]{Vucetic}
{Vu{\v{c}}eti{\'c}}, M.~M., {{\'C}iprijanovi{\'c}}, A., {Pavlovi{\'c}}, M.~Z.,
  {et~al.} 2015,
  \href{https://ui.adsabs.harvard.edu/abs/2015SerAJ.191...67V}{Serbian
  Astronomical Journal},
  \href{https://ui.adsabs.harvard.edu/abs/2015SerAJ.191...67V}{191, 67}

\end{thebibliography}
}

\begin{strip}

\end{strip}

{\ }

\clearpage

{\ }
\newpage

\begin{strip}

{\ }



\naslov{SOPSTVENO KRETANJE FILAMENATA U OSTATKU SUPERNOVE LABUDOVA PETLJA}

\authors{M. Vu{\v c}eti{\' c}$^{1}$, N. Milanovi{\' c}$^{2}$, D. Uro{\v s}evi{\' c}$^{1}$, J. Raymond$^{3}$, D. Oni{\' c}$^{1}$, S. Milo{\v s}evi{\' c}$^{1}$ and N. Petrov$^{4}$}

\vskip3mm


\address{$^1$Department of Astronomy, Faculty of Mathematics,
University of Belgrade\break Studentski trg 16, 11000 Belgrade,
Serbia}


\Email{milica.vucetic@matf.bg.ac.rs}

\address{$^2$Max Planck Institute for Solar System Research, Justus-von-Liebig-Weg 3, 37077 G{\"o}ttingen, Germany}

\address{$^3$Harvard-Smithsonian Center for Astrophysics, 60 Garden Street, Cambridge, MA, 02138, USA}

\address{$^4$Institute of Astronomy and National Astronomical Observatory, Bulgarian Academy of Sciences, Tsarigradsko Shose 72, BG-1784 Sofia, Bulgaria}


\centerline{{\rrm UDK} \udc}


\vskip1mm

\centerline{\rit Uredjivaqki prilog}

\vskip.7cm

\baselineskip=3.8truemm

\begin{multicols}{2}

{
\rrm

U ovom radu odredjene su brzine udarnog talasa u Galaktiqkom ostatku supernove Labudova petlja, mere{\cc}i sopstvena kretanja optiqkih filamenata u ovom ostatku. Sopstvena kretanja su odredjena poredjenjem  $\mathrm{H}\alpha$ snimaka ostatka u dve epohe. Prva epoha je bila 1993. godine (KPNO), a druga 2018. i 2019. godine (Nacionalna astronomska opservatorija Ro{\zz}en, Astronomska stanica Vidojevica). Brzina svakog filamenta je zatim odredjena kori{\ss}{\cc}enjem najnovije procene za udaljenost do Labudove petlje (725 $\pm$ 15 $\mathrm{pc}$; \cite{Fesen2021}). Odredili smo brzine udara za 35 pozicija du{\zz} razliqitih filamenata, {\ss}to je dvostruko vi{\ss}e nego u prethodnim studijama vezanim za  ovaj deo Labudove petlje. Po prvi put, izmerili smo brzine radijativnih filamenata u ovom regionu. Medju prouqavanim filamentima nalaze se tri merenja koja dolaze sa dva radijativna filamenta, koja su odabrana na osnovu toga {\ss}to su ti filamenti vidljivi na $\mathrm{[SII]}$ snimcima iz druge epohe. Brzine dobijene za neradijativne filamente su u opsegu 240--650  $\mathrm{km\ s^{-1}}$ sa gre{\ss}kom od 70  $\mathrm{km\ s^{-1}}$ {\ss}to je u saglasnosti sa prethodnim studijama. Tri radijativna filamenta imaju znatno ni{\zz}e brzine od 100--160$\pm$70 $\mathrm{km\ s^{-1}}$, {\ss}to je u saglasnosti sa pretpostavkom da su ti filamenti evolutivno stariji. Ovo dokazuje da se emisija u $\mathrm{[SII]}$ liniji mo{\zz}e koristiti za odabir radijativnih filamenata u ostacima supernovih.

{\ }

}

\end{multicols}

\end{strip}


\end{document}